\documentclass[twocolumn, prl,showpacs]{revtex4-1}

\usepackage{amsmath}
\usepackage{mathrsfs} 
\usepackage{amssymb} 
\usepackage{graphicx}
\usepackage{subfigure}
\usepackage{dcolumn}
\usepackage{bm}
\usepackage{natbib}

\begin{document}
\title{Strain-tunable topological quantum phase transition in buckled honeycomb lattices}
\date{\today}
\author{Jia-An Yan$^1$}
\email{jiaanyan@gmail.com}
\affiliation{1. Department of Physics, Astronomy, and Geosciences, Towson University, 8000 York Road, Towson, MD 21252, USA\\
2. Department of Physics, University of Arkansas, Fayetteville, Arkansas 72701, USA\\
3. Department of Physics, Washington University, St Louis, MO 63005, USA}
\author{Mack A.  Dela Cruz$^1$}
\affiliation{1. Department of Physics, Astronomy, and Geosciences, Towson University, 8000 York Road, Towson, MD 21252, USA\\
2. Department of Physics, University of Arkansas, Fayetteville, Arkansas 72701, USA\\
3. Department of Physics, Washington University, St Louis, MO 63005, USA}

\author{Salvador Barraza-Lopez$^2$}
\affiliation{1. Department of Physics, Astronomy, and Geosciences, Towson University, 8000 York Road, Towson, MD 21252, USA\\
2. Department of Physics, University of Arkansas, Fayetteville, Arkansas 72701, USA\\
3. Department of Physics, Washington University, St Louis, MO 63005, USA}

\author{Li Yang$^3$}
\affiliation{1. Department of Physics, Astronomy, and Geosciences, Towson University, 8000 York Road, Towson, MD 21252, USA\\
2. Department of Physics, University of Arkansas, Fayetteville, Arkansas 72701, USA\\
3. Department of Physics, Washington University, St Louis, MO 63005, USA}

\begin{abstract}
 Low-buckled silicene is a prototypical quantum spin Hall insulator with the topological quantum phase transition controlled by an out-of-plane electric field. We show that this field-induced electronic transition can be further tuned by an in-plane hydrostatic biaxial strain $\varepsilon$, owing to the curvature-dependent spin-orbit coupling (SOC): There is a $Z_2$ = 1 topological insulator phase for biaxial strain $|\varepsilon|$ smaller than 0.07, and the band gap can be tuned from 0.7 meV for $\varepsilon = +0.07$ up to a fourfold 3.0 meV for $\varepsilon = -0.07$. First-principles calculations also show that the critical field strength $E_c$ can be tuned by more than 113\%, with the absolute values nearly 10 times stronger than the theoretical predictions based on a tight-binding model. The buckling structure of the honeycomb lattice thus enhances the tunability of both the quantum phase transition and the SOC-induced band gap, which are crucial for the design of topological field-effect transistors based on two-dimensional materials.
\end{abstract}

\maketitle

Two-dimensional (2D) quantum spin Hall (QSH) insulator \cite{Kane2005v1,Kane2005v2,Bernevig2006,Bernevig2006sci,Konig2007,Hasan2010,Qi2011} is characterized by an insulating bulk and gapless edge states at its boundaries \cite{Kane2005v1,Kane2005v2}. These edge states are topologically protected from backscattering of non-magnetic defects or impurities due to time-reversal symmetry, thus providing enticing concepts for novel quantum electronic devices with low energy dissipation \cite{Hasan2010,Qi2011}. Quantized conductance through QSH edge states were originally reported on HgTe/CdTe \cite{Bernevig2006sci,Konig2007}, and on InAs/GaSb \cite{Liu2008,Du2013} quantum wells too.

\begin{figure}[tbp]
\centering
\includegraphics[width=8.5cm,clip]{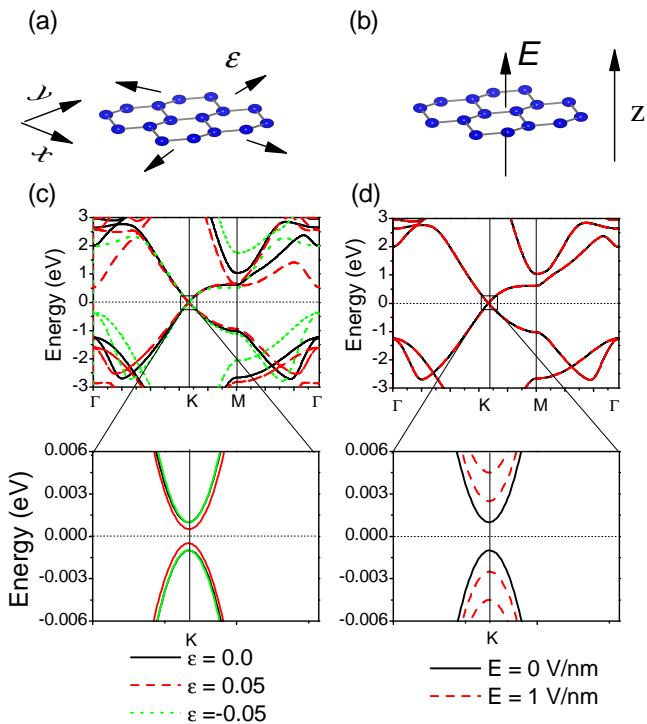}
 \caption{(Color online) The electronic properties of silicene are independently tuned by (a) an in-plane biaxial strain $\varepsilon$, and (b) an out-of-plane $E-$field $E_z$. Subplots (c) and (d) are band dispersions under typical values of $\varepsilon$ or $E_z$, respectively. Insets are zoom-ins of the band dispersion near the $K-$point (the Fermi level is set to zero; note the overlapping bands for $\varepsilon=0.00$ and $0.05$ on (c)). Here, we will explore the combined effects of $\varepsilon$ and $E_z$ on its electronic structures.}\label{fig1}
\end{figure}

There is an intense drive to realize QSH insulators with controllable quantum phase transitions and tunable electronic and spin properties \cite{Ezawa2012njp,Ezawa2012prl}. The intrinsic spin-orbit coupling (SOC) in graphene is weak \cite{Kane2005v2,Min2006,HH2006,Yao2007}, but other 2D materials that may realize this phenomena include honeycomb lattices of bismuth atoms on a silicon surface \cite{FengPNAS}, an electric-field-induced QSH phase on few-layer black phosphorus \cite{Liu2014}, a new structural phase of a transition-metal dichalcogenide \cite{Guinea2014,Qian2014}, and silicene/germanene  \cite{Ciraci,Liu2011prl,Liu2011prb,Ni2012,Drummond2012,Ezawa2012njp,Ezawa2012prl,Padilha2013}, among others \cite{Ma,PRL2013,Rapid2014,Tang2014}.

Mechanical strain and curvature (i.e., {\em shape}) \cite{HH2006,Maria2010} modify the electronic and spin properties of buckled honeycomb lattices, and a deep understanding of their effects is emerging \cite{ACSNano,RapidAlejandro,Tomanek,Kamien}. In a classic work, the dependence of the SOC strength $\lambda_{SOC}$ on curvature is established to arise from (1) on-site spin-flips among $\sigma-$ and $\pi-$ bands due to the intrinsic coupling among spin and $s-$ and $p-$electronic orbitals, and (2) a subsequent hybridization of the hopping $\pi-$electron with electrons from bands with $\sigma-$symmetry in the presence of curvature \cite{HH2006}. This important phenomenon lies beyond those descriptions of the electronic structure of silicene and other group-IV 2D materials that are based on $\pi-$electrons only \cite{Ezawa2012njp,Ezawa2012prl}. Curvature is of central importance for the discussion of 2D-based topological field-effect transistors because it raises the intrinsic gap, and hence the temperature at which these devices could operate. On the other hand, a controllable and tunable topological quantum phase transition will greatly facilitate the device fabrication and operation \cite{Qian2014}. Working on silicene, we generalize Ezawa's result $\Delta(E_z)=2|l E_z-\lambda_{SOC}|$ for the dependence of the energy band gap $\Delta(E_z)$ on the electric field ($E-$field) $E_z$, in which the SOC strength $\lambda_{SOC}$ is assumed to be a constant, into:
\begin{equation}
\Delta(E_z,\varepsilon)=2|l(\varepsilon)E_z-\lambda_{SOC}(\varepsilon)|,
\end{equation}
where $\lambda_{SOC}$ evolves with strain due to a curvature-induced hybridization among $s$ and $p$ electrons.

Pristine silicene is a QSH insulator with a band gap $\Delta(E_z=0,\varepsilon=0)$ of about 1.5 meV (18 K) \cite{Liu2011prl,Liu2011prb}. This band gap is tunable by an $E-$field $E_z$ perpendicular to the buckled atomic layer \cite{Ni2012,Drummond2012,Ezawa2012njp} as an electrostatic potential difference is established between the two Si atoms in the unit cell due to their height difference $2l$. Silicene behaves as a $Z_2$ = 1 non-trivial QSH insulator below a critical $E_z$ strength $|E_c|$, and becomes a trivial band insulator for values of $|E_z|>|E_c|$ \cite{Ezawa2012njp,Ezawa2012prl}. In addition, silicene is likely to be fabricated on substrates \cite{Aufray2010,Lalmi2010,Vogt2012,Fleurence2012,Chen2012,Feng2012,Jamgotchian2012,Lin2012} and may be subject of in-plane strain $\varepsilon$ already. A fundamental understanding of effects of the strain on the topological quantum phase transition is important for the design of quantum electronic devices based on silicene and other 2D materials \cite{FengPNAS,Liu2014,Qian2014,Padilha2013}.

Calculations were carried out using density-functional theory \cite{HK1964,KS1965} as implemented in the Quantum ESPRESSO code \cite{pwscf} with the GGA-PBEsol exchange-correlation functional \cite{PBE2008}. The SOC was included in the fully relativistic Rappe-Rabe-Kaxiras-Joannopoulos (RRKJ) ultrasoft pseudopotential scheme \cite{RRKJ1990} with a nonlinear core correction. The cutoff energy in the plane wave expansion is 50 Ry. A Monkhorst-Pack uniform $k$-grid of 36$\times$36$\times$1 is employed. A vacuum region of 20 \AA~ is introduced along the out-of-plane ($z$) direction to eliminate spurious interactions among periodic images. The $E-$field $E_z$ is induced by a saw-tooth potential along the $z-$direction (c.f., Fig.~1(b)). Biaxial strain is applied to the silicene lattice, and $\varepsilon$ is defined as $\varepsilon$ = $(a-a_0)/a_0$ (see Fig.~1(a)). Here $a$ and $a_0=3.85$ \AA{} \cite{Bechstedt2012,Padilha2013,Rapid2014} are the strained and unstrained lattice constants, respectively.

\begin{figure}[tbp]
\centering
\includegraphics[width=8.5cm,clip]{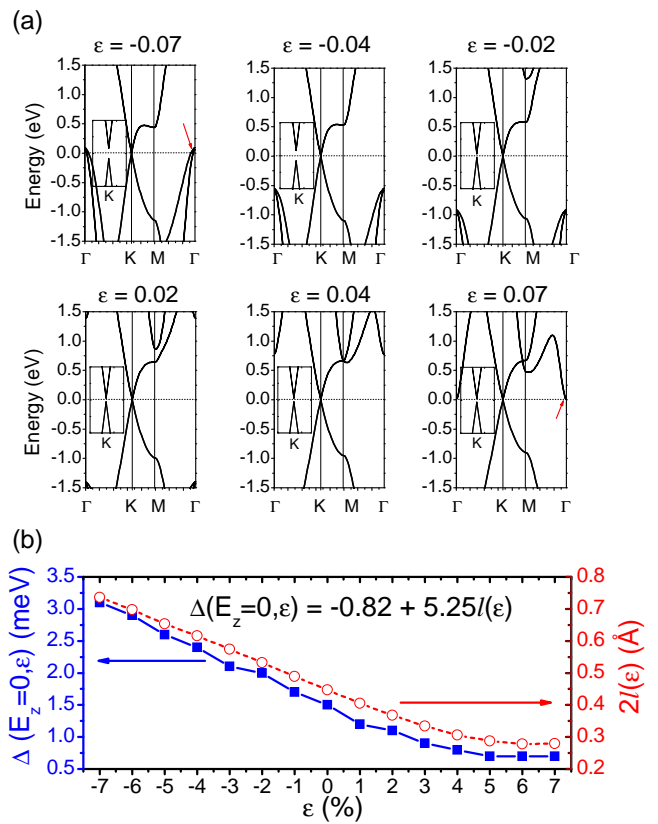}
 \caption{(a) Typical band dispersions for various magnitudes of $\varepsilon$: A semicondoctor-to-metal transition occurs when $|\varepsilon|> 0.07$, as highlighted by the tilted (red) arrows. Insets help highlight the band dispersions near the Fermi level around the $K-$point. (b) Band gap $\Delta$ (left axis) and the height difference $2l$ due to buckling (right axis) as a function of biaxial strain $\varepsilon$, for an $E_z=0$ display a linear relationship in silicene, that has been explicitly indicated.  }\label{fig2}
\end{figure}

There is a 1.5 meV gap opening at the $K-$point that is induced by the SOC for $\varepsilon=0$ and $E_z=0$. When independently applied (c.f., Figs.~1(a) and 1(b)) strain and $E-$field lead to distinct changes on the electronic band dispersions, as seen in Figs.~1(c) and 1(d), respectively. In-plane strain preserves inversion symmetry, leading to a gap opening that is more apparent for compressive strain (see the inset of Fig.~1(c), $\varepsilon=-0.05$), thus suggesting a relation among $\lambda_{SOC}$ and $\varepsilon$ that --as a matter of fact-- has been discussed in the context of graphene some time ago \cite{HH2006}. On the other hand, $E_z$ lifts inversion symmetry thus removing the band degeneracy, as seen on the inset of Fig.~1(d) for $E_z=1$ V/nm (dashed lines). The effects of $E-$fields on the electronic properties of silicene have been thoroughly discussed in the past, so we continue exploring the effect of strain on Fig.~\ref{fig2}.

There are three effects of strain on the electronic structure that are conveyed by Fig.~\ref{fig2}(a): (i) A renormalization of the Fermi velocity near the $K-$point \cite{deJuan}. (ii) An upward shift of the valence-band maxima (VBM) under compressive strain, and a downward shift of the conduction-band minima (CBM) under tensile strain at the $\Gamma-$point. (iii) A gap opening at the $K-$point that is especially evident under compressive strain. Effects (ii) and (iii) are responsible for the semiconductor-to-metal transition, and are the focus of the ensuing discussion.

For sufficiently large compressive strain, the VBM crosses the Fermi level at the $\Gamma-$point (c.f., Fig.~\ref{fig2}(a) for $\varepsilon=-0.07$; this crossing has been highlighted by a tilted red arrow). Similarly, the CBM at the $\Gamma-$point nearly touches the Fermi level for a tensile strain of $\varepsilon=+$0.07: Silicene undergoes a transition to a metallic phase for strain beyond $|\varepsilon|\sim0.07$. Although arising under different mechanisms, these effects due to {\em strain} are reminiscent of the topological-metal phase transition that is induced by an {\em electric field} in few-layer black phosphorus \cite{Liu2014}.

\begin{figure}[tbp]
\centering
\includegraphics[width=8.5cm,clip]{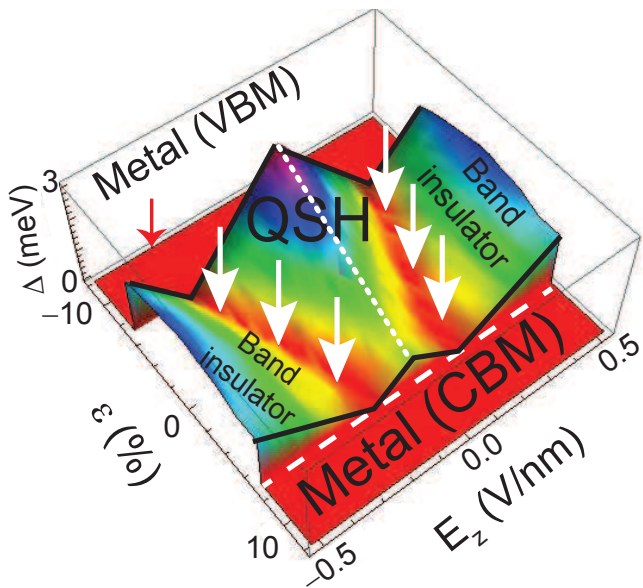}
 \caption{(Color online) Topological quantum phase diagram of silicene with respect to in-plane biaxial strain $\varepsilon$ and out-of-plane $E-$field $E_z$. The vertical axis is the band gap $\Delta$. The critical $E-$fields $E_c$, at which there is a phase transition from a topological insulator into a band insulator, have been indicated by vertical white arrows. The metallic state has a zero value of $\Delta$ that is shown in red, and the area marked by QSH represents the topological spin Hall state phase. }\label{fig3}
\end{figure}

We next discuss the band dispersion seen around the $K-$point in the insets of Fig.~\ref{fig2}(a). To do so, we first report the SOC-induced band gap $\Delta$ and the buckling height $2l$ as a function of $\varepsilon$ in Fig.~\ref{fig2}(b): $\Delta$ increases monotonically as $\varepsilon$ decreases from $\varepsilon=+0.07$ down to $\varepsilon=-0.07$, meaning that {\em compressive strain enhances the band gap $\Delta$} while tensile strain decreases the gap. Since the energy band gap never closes in going from pristine silicene -- a $Z_2 = 1$ topological insulator -- to strained silicene with $|\varepsilon| < 0.07$, so these systems share the same topological classification according to the adiabatic continuity argument for transformations of the electronic bands.

The above phenomenon can be understood from the change of the buckling height $2l$ among two silicon atoms due to an in-plane strain ($l=l(\varepsilon)$) shown in Fig.~\ref{fig2}(b). Buckling (and hence $l$) increases with compressive strain, thus enhancing the overlap among $\sigma$ and $\pi$ orbitals, resulting in a strain-dependent SOC \cite{HH2006,Liu2011prl}. The band gap $\Delta$ increase is almost linear on $l$: $\Delta \sim 5.25 l$. A positive $\varepsilon$ reduces $l$, making the structure more planar (graphene-like), so the overlap among $\sigma$ and $\pi$ orbitals decreases, bringing $\lambda_{SOC}$ down into its ``intrinsic value'' (i.e., its value under a zero curvature) \cite{HH2006,Liu2011prl}. However, when $\varepsilon$ = $+$0.07, the CBM at the $\Gamma-$point nearly touches the Fermi level (tilted arrow on subplot in Fig.~2(a)), so that a further increase of $\varepsilon$ induces a transition into a metallic phase.

The strain-dependent SOC may be more qualitatively explained with the following matrix form at atom $A$ \cite{Liu2011prb}:
\begin{equation}\label{eq:soc}
H_{A,SOC}=\frac{\tilde{\lambda}_{SOC}}{2}\left(
\begin{matrix}
$0$ &$0$  & $0$  & $0$ &$0$ &$0$ & $0$ & $0$\\
$0$ &$0$  &$-i$& $0$ &$0$ &$0$ & $0$ & $1$\\
$0$ &$+i$ & $0$  & $0$ &$0$ &$0$ & $0$ & $-i$\\
$0$ &$0$  & $0$  & $0$ &$0$ &$1$ & $-i$ & $0$\\
$0$ &$0$  & $0$   &$0$   &$0$ &$0$ &$0$    & $0$\\
$0$ &$0$  & $0$   &$1$   &$0$ &$0$ & $-i$ & $0$\\
$0$ &$0$  & $0$   &$+i$  &$0$ &$+i$ &$0$ & $0$\\
$0$ &$1$  & $+i$  &$0$   &$0$ &$0$ &$0$ &$0$
\end{matrix}
\right),
\end{equation}
on the basis set \{$|s\uparrow\rangle$, $|p_x\uparrow\rangle$, $|p_y\uparrow\rangle$, $|p_z\uparrow\rangle$,  $|s\downarrow\rangle$, $|p_x\downarrow\rangle$, $|p_y\downarrow\rangle$, $|p_z\downarrow\rangle$\}. This SOC matrix produces spin flips among $|p_z\rangle$ ($\pi$) and $|p_x\rangle$, $|p_y\rangle$ orbitals (that belong on $\sigma-$ bands prior to the orbital hybridization) and $\tilde{\lambda}_{SOC}$ could be obtained as a fitting parameter. The SOC matrix at atom $B$ $H_{B,SOC}$ has an identical form. This {\em intrinsic} SOC is further enhanced by hopping~\cite{HH2006}.

After indicating the dependence among $\Delta$ and $\varepsilon$ seen on the first-principles data (Figs.~1 and 2), we study the combined effects of strain and $E-$field on the topological phase transition, and display $\Delta$ as a function of the $E_z$ and $\varepsilon$.  $\Delta(E_z,\varepsilon)$ has a characteristic $W-$shape \cite{Liu2011prl,Liu2011prb,Ezawa2012njp,Padilha2013,Qian2014} as seen in Fig.~3. For a given value of $\varepsilon$, $\Delta$ decreases to zero as $|E_z|$ increases. For a given magnitude of $\varepsilon$ smaller than 7\%, the value of $E_z$ for which $\Delta=0$ is known as the critical field $|E_c(\varepsilon)|$: As the field increases further beyond $|E_c(\varepsilon)|$, the energy band gap $\Delta$ reopens and increases again, but the electronic state is a trivial insulator because, according to the bulk-boundary correspondence principle, the topological phase transition occurs at the critical field strength $|E_c(\varepsilon)|$ when the band gap closes. Fig.~\ref{fig3} clearly shows that the critical field strength $|E_c(\varepsilon)|$ increases with compressive strain: {\em Strain tunes the quantum phase transition due to a buckling-dependent SOC.}

Fig.~\ref{fig3} can also be seen as a phase diagram that indicates the different quantum phases that can be reached by jointly tuning $E_z$ and $\varepsilon$: Between $-$0.07 $<\varepsilon<$+0.07, silicene is a QSH insulator for any applied $E-$field in the range of $|E|<|E_c(\varepsilon)|$. On the other hand, silicene changes to an ordinary band insulator with $|E|>|E_c(\varepsilon)|$. This phase diagram is clearly visible in Fig.~\ref{fig3}.

\begin{figure}[tbp]
\centering
\includegraphics[width=8.5cm,clip]{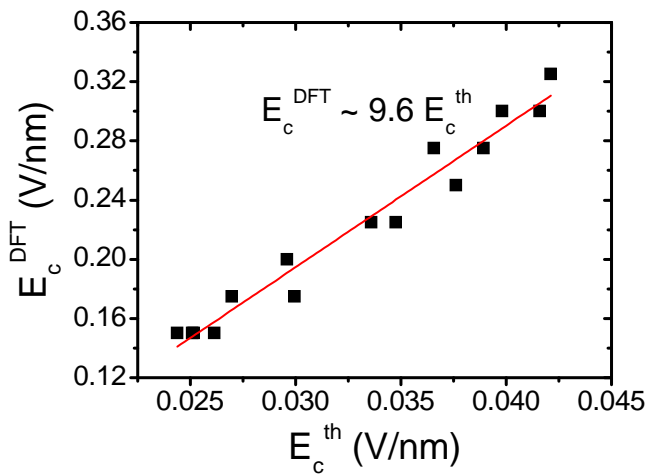}
 \caption{(Color online) The critical $E-$field strength $E_c^{\texttt{DFT}}$ calculated from DFT is shown as a function of the theoretical value $E_c^{\texttt{th}}$ based on Eq.~(1). The red line indicates a linear fit of the data. }\label{fig4}
\end{figure}

Finally, we discuss the critical field strength $E_c$. In Fig.~\ref{fig4}, the first-principles data $E_c^{\texttt{DFT}}$ are shown as a function of the theoretical predictions from Eq.~(1): $E_c^{\texttt{th}} = \lambda_{SOC}/l$. The DFT results are nearly 10 times larger than the theoretical values, indicating a strong screening in silicene. In Ref. [\onlinecite{Drummond2012}], the field-induced band gap is found to be suppressed by a factor of about eight due to the high polarizability of silicene. Here, we show that the screening also significantly enhances the critical $E-$field strength required to induce a quantum phase transition. More precisely, a linear fit in Fig.~4 yields $E_c^{\texttt{DFT}} \sim 9.6 E_c^{\texttt{th}}$.

In summary, using first-principles methods, we show that the biaxial strain $\varepsilon$ can be utilized to tune the spin-orbit coupling in silicene and hence its topological quantum phase transition. At $|\varepsilon|$$\sim$0.07, silicene undergoes a transition from topological insulator into a metallic phase. Within the range of $-$0.07 $<$$\varepsilon$$<+$0.07, pristine silicene remains a QSH insulator with a strain-dependent SOC that increases under a compressive strain. The critical electric field strength is significantly enhanced by the screening, nearly 10 times larger than the theoretical predictions from a tight-binding model. These phenomena highlight the interplay between the mechanical strain and $E-$field on the electronic properties of low-buckled honeycomb lattices.

J.A.Y. acknowledges the Faculty Development and Research Committee grant (OSPR No. 140269) and the FCSM Fisher General Endowment at the Towson University. M.A.D.C. is partially supported by FCSM Fisher General Endowment at the Towson University. S.B.L. Acknowledges funding from the Arkansas Biosciences Institute and NSF-XSEDE (TG-PHY090002).


\begin{thebibliography}{99}

\bibitem{Kane2005v1} C. L. Kane and E. J. Mele, Phys. Rev. Lett. \textbf{95}, 226801 (2005).

\bibitem{Kane2005v2} C. L. Kane, E. J. Mele, Phys. Rev. Lett. \textbf{95}, 146802 (2005).

\bibitem{Bernevig2006} B. A. Bernevig, S.-C. Zhang, Phys. Rev. Lett. \textbf{96}, 106802 (2006).

\bibitem{Bernevig2006sci} B. A. Bernevig, T. L. Hughes, S.-C. Zhang, Science \textbf{314},1757 (2006).

\bibitem{Konig2007} M. K\"{o}nig, S. Wiedmann, C. Br\"{u}ne, A. Roth, H. Buhmann, L. W. Molenkamp, X.-L. Qi,
 and S.-C. Zhang, Science \textbf{318}, 766 (2007).

\bibitem{Hasan2010} M. Z. Hasan, C. L. Kane, Rev. Mod. Phys. \textbf{82}, 3045 (2010).

\bibitem{Qi2011} X.-L. Qi, S.-C. Zhang, Rev. Mod. Phys. \textbf{83}, 1057 (2011).

\bibitem{Liu2008} C. Liu, T. L. Hughes, X. L. Qi, K. Wang, S. C. Zhang, Phys. Rev. Lett. \textbf{100}, 236601 (2008).

\bibitem{Du2013} L. Du, I. Knez, G. Sullivan, R.-R. Du, http://arxiv.org/abs/1306.1925.


\bibitem{Ezawa2012njp} M. Ezawa, New J. Phys. \textbf{14}, 033003 (2012).

\bibitem{Ezawa2012prl} M. Ezawa, Phys. Rev. Lett. \textbf{109}, 055502 (2012).

\bibitem{Min2006} H. Min, J. E. Hill, N. A. Sinitsyn, B. R. Sahu, L. Kleinman, and A. H. MacDonald, Phys. Rev. B \textbf{74}, 165310 (2006).

\bibitem{HH2006} D. Huertas-Hernando, F. Guinea, and A. Brataas, Phys. Rev. B \textbf{74}, 155426 (2006).

\bibitem{Yao2007} Y. Yao, F. Ye, X.-L. Qi, S.-C. Zhang, and Z. Fang, Phys. Rev. B \textbf{75}, 041401(R) (2007).


\bibitem{FengPNAS} M. Zhou, W. Ming, Z. Liu, Z. Wang, P. Li, and F. Liu, Proc. Natl. Acad. Sci. (USA) \textbf{111}, 14378 (2014).

\bibitem{Liu2014} Q. Liu, X. Zhang, L. B. Abdalla, A. Fazzio, and A. Zunger, http://arxiv.org/abs/1411.3932.

\bibitem{Guinea2014}M.~A. Cazalilla, H. Ochoa, and F. Guinea. Phys. Rev. Lett. {\bf 113}, 077201 (2014).

\bibitem{Qian2014} X. Qian, J. Liu, L. Fu, and J. Li, Science \textbf{346}, 1344 (2014).

\bibitem{Ciraci} S. Cahangirov, M. Topsakal, E. Akt\"{u}rk, H. Sahin, and S. Ciraci, Phys. Rev. Lett. \textbf{102}, 236804 (2009).

\bibitem{Liu2011prl} C. C. Liu, W. X. Feng, and Y. G. Yao, Phys. Rev. Lett. \textbf{107}, 076802 (2011).

\bibitem{Liu2011prb} C. C. Liu, H. Jiang, and Y. G. Yao, Phys. Rev. B \textbf{84}, 195430 (2011).


\bibitem{Ni2012} Z. Ni, Q. Liu, K. Tang, J. Zheng, J. Zhou, R. Qin, Z. Gao,
D. Yu, and J. Lu, Nano Lett. \textbf{12}, 113-118 (2012).

\bibitem{Drummond2012} N. D. Drummond, V. Z\'{o}lyomi, and V. I. Fal'ko, Phys. Rev. B \textbf{85}, 075423 (2012).



\bibitem{Padilha2013} J. E. Padilha, L. Seixas, R. B. Pontes, A. J. R. da Silva, and A. Fazzio, Phys. Rev. B \textbf{88}, 201106(R) (2013).

\bibitem{Ma} Y. Ma, Y. Dai, M. Guo, C. Niu, and B. Huang, J. Phys. Chem. C \textbf{116}, 12977 (2012).

\bibitem{PRL2013}Y. Xu, B. Yan, H.-J. Zhang, J. Wang, G. Xu, P. Tang, W. Duan, and S.-C. Zhang, Phys. Rev. Lett. \textbf{111}, 136804 (2013).

\bibitem{Rapid2014} P. Rivero, J.-A. Yan, V. M. Garcia-Suarez, J. Ferrer, and S. Barraza-Lopez. Phys. Rev. B \textbf{90}, 241408(R) (2014).

\bibitem{Tang2014}P. Tang, P. Chen, W. Cao, H. Huang, S. Cahangirov, L. Xian, Y. Xu, S.-C. Zhang, W. Duan, and A. Rubio. Phys. Rev. B \textbf{90}, 121408(R) (2014).

\bibitem{Maria2010}M. A. H. Vozmediano, M. I. Katsnelson, and F. Guinea. Phys. Rep. {\bf 496}, 109 (2010).

\bibitem{ACSNano}A. A. Pacheco Sanjuan, Z. Wang, H. Pour Imani, M. Vanevic, and S. Barraza-Lopez. Phys. Rev. B {\bf 89}, 121403(R) (2014).
    %
\bibitem{RapidAlejandro}A. A. Pacheco Sanjuan, M. Mehboudi, E. O. Harriss, H. Terrones, and S. Barraza-Lopez. ACS Nano {\bf 8}, 1136 (2014).

\bibitem{Tomanek} J. Guan, Z. Jin, Z. Zhu, C. Chuang, B.-Y. Jin, and D. Tom{\'a}nek. Phys. Rev. B {\bf 90}, 245403 (2014).

\bibitem{Kamien}T. Castle, Y. Cho, X. Gong, E. Jung, D. M. Sussman, S. Yang, and R. D. Kamien. Phys. Rev. Lett. {\bf 113}, 245502 (2014).

\bibitem{deJuan} F. de Juan, M. Sturla, and M. A. H. Vozmediano. Phys. Rev. Lett. {\bf 108}, 227205 (2012).

\bibitem{Aufray2010} B. Aufray, A. Kara, S. Vizzini, H. Oughaddou, C. Leandri, B. Ealet, and G. Le Lay, Appl. Phys. Lett. \textbf{96}, 183102 (2010).

\bibitem{Lalmi2010} B. Lalmi, H. Oughaddou, H. Enriquez, A. Kara, S. Vizzini, B. Ealet, and B. Aufray, Appl. Phys. Lett. \textbf{97}, 223109 (2010).


\bibitem{Vogt2012} P. Vogt, P. DePadova, C. Quaresima, J. Avila, E. Frantzeskakis, M. C. Asensio, A. Resta, B. Ealet,
and G. LeLay, Phys. Rev. Lett. \textbf{108}, 155501 (2012).

\bibitem{Fleurence2012} A. Fleurence, R. Friedlein, T. Ozaki, H. Kawai, Y. Wang, and Y. Yamada-Takamura, Phys. Rev. Lett. \textbf{108}, 245501 (2012).

\bibitem{Chen2012} L. Chen, C.-C. Liu, B. Feng, X. He, P. Cheng, Z. Ding, S. Meng, Y. Yao, K. Wu, Phys. Rev. Lett. \textbf{109}, 056804 (2012).

\bibitem{Feng2012} B. Feng, Z. Ding, S. Meng, Y. Yao, X. He, P. Cheng, L. Chen, and K. Wu, Nano Lett., \textbf{12}, 3507-3511 (2012).

\bibitem{Jamgotchian2012} H. Jamgotchian, Y. Colignon, N. Hamzaoui, B. Ealet, J. Y. Hoarau, B. Aufray, and J. P. Bib\'{e}rian, J. Phys. Condens. Matter \textbf{24}, 172001 (2012).

\bibitem{Lin2012} C.-L. Lin, R. Arafune, K. Kawahara, N. Tsukahara, E. Minamitani, Y. Kim, N. Takagi, and M. Kawai, App. Phys. Exp. \textbf{5}, 045802 (2012).

\bibitem{HK1964} P. Hohenberg and W. Kohn, Phys. Rev. \textbf{136}, B864 (1964).
\bibitem{KS1965} W. Kohn and L. J. Sham, Phys. Rev. \textbf{140}, A1133 (1965).

\bibitem{pwscf} P. Giannozzi, S. Baroni, N. Bonini, M. Calandra, R. Car, C. Cavazzoni, D. Ceresoli, G. L. Chiarotti, M. Cococcioni, I. Dabo, A. Dal Corso, S. Fabris, G. Fratesi, S. de Gironcoli, R. Gebauer, U. Gerstmann, C. Gougoussis, A. Kokalj, M. Lazzeri, L. Martin-Samos, N. Marzari, F. Mauri, R. Mazzarello, S. Paolini, A. Pasquarello, L. Paulatto, C. Sbraccia, S. Scandolo, G. Sclauzero, A. P. Seitsonen, A. Smogunov, P. Umari, R. M. Wentzcovitch, J. Phys. Condens. Matter \textbf{21}, 395502 (2009).

\bibitem{PBE2008} J. P. Perdew, A. Ruzsinszky, G. I. Csonka, O. A. Vydrov, G. E. Scuseria, L. A. Constantin, X. Zhou, and K. Burke, Phys. Rev. Lett. \textbf{100}, 136406 (2008); Erratum Phys. Rev. Lett. \textbf{102}, 039902(E) (2009).

\bibitem{RRKJ1990} A.~M. Rappe, K.~M. Rabe, E. Kaxiras, and J.~D. Joannopoulos, Phys. Rev. B \textbf{41}, 1227(R) (1990); Erratum Phys. Rev. B 44, 13175 (1991).
%

\bibitem{Bechstedt2012} F. Bechstedt, L. Matthes, P. Gori, and O. Pulci, App. Phys. Lett. \textbf{100}, 261906-261908 (2012).

\end{thebibliography}
\end{document}